\documentclass[twocolumn,footinbib,preprintnumbers,floatfix,aps,prl,10pt]{revtex4-2}
\usepackage{amsmath,amssymb,graphicx,bm,psfrag,color,slashed,mathtools}
\usepackage{hyperref}
\usepackage[dvipsnames]{xcolor}

\newcommand{\spac}{{\hspace{0.3mm}}}

\makeatletter
\DeclareFontFamily{OMX}{MnSymbolE}{}
\DeclareSymbolFont{MnLargeSymbols}{OMX}{MnSymbolE}{m}{n}
\SetSymbolFont{MnLargeSymbols}{bold}{OMX}{MnSymbolE}{b}{n}
\DeclareFontShape{OMX}{MnSymbolE}{m}{n}{
    <-6>  MnSymbolE5
   <6-7>  MnSymbolE6
   <7-8>  MnSymbolE7
   <8-9>  MnSymbolE8
   <9-10> MnSymbolE9
  <10-12> MnSymbolE10
  <12->   MnSymbolE12
}{}
\DeclareFontShape{OMX}{MnSymbolE}{b}{n}{
    <-6>  MnSymbolE-Bold5
   <6-7>  MnSymbolE-Bold6
   <7-8>  MnSymbolE-Bold7
   <8-9>  MnSymbolE-Bold8
   <9-10> MnSymbolE-Bold9
  <10-12> MnSymbolE-Bold10
  <12->   MnSymbolE-Bold12
}{}

\let\llangle\@undefined
\let\rrangle\@undefined
\DeclareMathDelimiter{\llangle}{\mathopen}%
                     {MnLargeSymbols}{'164}{MnLargeSymbols}{'164}
\DeclareMathDelimiter{\rrangle}{\mathclose}%
                     {MnLargeSymbols}{'171}{MnLargeSymbols}{'171}
\makeatother

\makeatletter 
    
\renewcommand\onecolumngrid{
\do@columngrid{one}{\@ne}%
\def\set@footnotewidth{\onecolumngrid}
\def\footnoterule{\kern-6pt\hrule width 1.5in\kern6pt}%
}

\renewcommand\twocolumngrid{
        \def\footnoterule{
        \dimen@\skip\footins\divide\dimen@\thr@@
        \kern-\dimen@\hrule width.5in\kern\dimen@}
        \do@columngrid{mlt}{\tw@}
}%

\makeatother    

\begin{document}
\preprint{CERN-TH-2026-039}
\preprint{MITP-26-008}
\date{March 12, 2026}

\title{Factorization Beyond Coherence}

\author{Thomas Becher$^a$}
\author{Patrick Hager$^b$}
\author{Matthias Neubert$^{c,d}$}
\author{Dominik Schwienbacher$^a$}
\affiliation{${}^a$Institut f\"ur Theoretische Physik {\em \&} AEC, Universit\"at Bern, Sidlerstrasse 5, CH-3012 Bern, Switzerland\\
${}^b$CERN, Theoretical Physics Department, CH-1211 Geneva 23, Switzerland\\
${}^c$PRISMA$^{++}$ Cluster of Excellence {\em \&} MITP, Johannes Gutenberg University, 55099 Mainz, Germany\\
${}^d$Department of Physics, LEPP, Cornell University, Ithaca, NY 14853, U.S.A.}

\begin{abstract}
\vspace{-3mm} 
We derive a novel factorization theorem for $N$-jettiness at hadron colliders, which incorporates coherence-violating effects induced by Glauber gluons and several new momentum modes. Their interplay generates coherence-violating logarithms (CVLs) starting at four-loop ($N\ge1$) or five-loop order ($N=0$). We calculate the anomalous dimensions required for the resummation of CVLs and establish a general framework for the systematic treatment of coherence violation. Our findings imply that most existing factorization formulas for global LHC observables must be revised.
\end{abstract}

\maketitle

Color coherence denotes the phenomenon that soft gluon radiation in hard scattering processes is sensitive only to the net color charge of a group of collinear partons rather than their individual charges. Together with soft-collinear factorization, it is a foundational principle that underpins precision collider phenomenology. They provide a systematic framework through which short-distance partonic dynamics can be disentangled from universal long-distance hadronic physics, thereby enabling systematically improvable predictions for jet observables. 
As the LHC enters the high-luminosity era, with percent-level experimental precision becoming the norm, this paradigm is pushed to its limits. Reliable phenomenology now calls for the inclusion of higher-order perturbative corrections and the precision resummation of large logarithmic corrections. 

Global jet observables such as $N$-jettiness~\cite{Stewart:2010tn}, which impose constraints on radiation across the full phase space, are particularly simple and play a central role in this program. They are useful in the context of slicing schemes~\cite{Gaunt:2015pea,Boughezal:2015dva,Boughezal:2015ded,Boughezal:2015aha,Campbell:2019gmd}, as well as for jet merging in parton showers~\cite{Alioli:2012fc,Alioli:2023rxx}. Their theoretical description relies on factorization formulas assuming a strict separation between collinear and soft modes, enabling all-order resummation of logarithmically-enhanced contributions.

Yet, it is known that strict collinear factorization at the amplitude level already fails at one-loop order for space-like splittings~\cite{Catani:2011st}. While this effect cancels in observables at the cross-section level, at higher loop orders~\cite{Schwartz:2017nmr,Henn:2024qjq,Guan:2024hlf} this is generally no longer the case~\cite{Catani:2011st,Forshaw:2012bi}.
The resulting breakdown manifests itself through coherence-violating effects, most prominently in the form of super-leading logarithms (SLLs) in non-global observables~\cite{Forshaw:2008cq,Becher:2021zkk,Becher:2023mtx,Dasgupta:2025cgl}. These SLLs are double-logarithmic effects in otherwise single-logarithmic observables at hadron colliders, starting at four-loop order. They originate from the non-cancellation of Glauber phases, arising from Coulomb-like gluon exchanges between soft and collinear sectors, thereby spoiling factorization.

Since SLLs arise in non-global observables with complicated phase-space constraints, it was hoped that for the simpler global observables these effects would be absent. However, this expectation has come under increasing scrutiny:\ coherence-violating logarithms (CVLs), of which SLLs are a particular subset, have been argued to arise even for global observables~\cite{Banfi:2010xy,Gaunt:2014ska,Forshaw:2021fxs}, and a recent analysis~\cite{Banfi:2025mra} has identified the leading-order CVL for $1$-jettiness using Lund-plane arguments. These developments challenge the  assumption that globalness guarantees soft-collinear factorization and raise a pressing question:\ do standard factorization formulas remain valid for even the simplest observables at hadron colliders?

In this Letter, we address this question and derive a novel factorization theorem for $N$-jettiness that accounts for the subtleties arising from genuine Glauber gluons in active-active parton scattering processes~\cite{Becher:2024kmk,Becher:2025igg}. We identify several new modes that arise only through interactions with Glauber gluons, leading to a much richer and more intricate structure of global observables than previously assumed, exhibiting features characteristic of their non-global counterparts. In addition, we present the one-loop anomalous dimensions sufficient for the all-order resummation of the leading CVLs.

We begin our analysis with the definition of the $N$-jettiness observable at hadron colliders, which reads~\cite{Stewart:2010tn}
\begin{equation}
   \mathcal{T}_N = 2\sum_k \min\bigg\{ \frac{q_a\cdot p_k}{Q_a},\frac{q_b\cdot p_k}{Q_b},\frac{q_1\cdot p_k}{Q_1},\dots,\frac{q_{N}\cdot p_k}{Q_N} \bigg\} \,,
\end{equation}
where $k$ runs over the final-state particles with momenta $p_k$. The light-like reference vectors $q_{a,b}$ and $q_i$ (with $i=1,\dots,N$) are defined as
\begin{equation}
   q_{a,b}^\mu = x_{a,b} \frac{\sqrt{s}}{2}\spac n_{a,b}^\mu \,, \quad 
   q_i^\mu = E_i\spac n_i^\mu \,.
\end{equation}
The center-of-mass energy is denoted by $s$. The definition of the hard scales $Q_i$ is not unique~\cite{Jouttenus:2013hs}, a common choice is $Q_i=2E_i$ and $Q_{a,b} = x_{a,b} \sqrt{s}$, where $x_{a,b}$ are the momentum fractions of the incoming partons. In the following, we abbreviate $n_a\equiv n$, $n_b \equiv \bar{n}$, with $n\cdot \bar{n}=2$. For $\mathcal{T}_N\equiv\lambda^2\spac Q\ll Q$, where $\lambda$ is a small expansion parameter and $Q$ is a reference hard scale. The final state consists of $N$ narrow jets, two beam jets, and ultra-soft radiation between the jets. Using soft collinear effective theory (SCET)~\cite{Bauer:2001yt,Bauer:2002nz,Beneke:2002ph,Beneke:2002ni}, the well-known factorization formula for $N$-jettiness takes the  form~\cite{Stewart:2010tn}
\begin{equation}\label{eq:fact_old}
   \frac{d\sigma}{d\mathcal{T}_N} = \Big\langle \bm{\mathcal{H}}_{N}\otimes B_a\otimes B_b\otimes \Bigl[ \prod_{i=1}^N J_{i} \Bigr]\otimes \bm{S}_{N} \Big\rangle \,,
\end{equation}
where $\bm{\mathcal{H}}_{N}$ denotes the hard function for $N$ outgoing partons, $B_{a,b}$ are the  beam functions for the incoming partons, $J_{i}$ is the jet function associated with direction $n_i$, and $ \bm{S}_{N}$ represents the ultra-soft function. The sum over partonic channels is kept implicit in this notation. The angular brackets denote the trace in color-helicity space of the partons. The boldface notation indicates operators acting in color space~\cite{Catani:1996vz}. The convolutions in the $\mathcal{T}_N$ variables of the various sectors are contained schematically in the  $\otimes$ symbol. The measurement functions are included in the definitions of the beam, jet and ultra-soft functions, with the PDFs contained within the beam functions. Details can be found in~\cite{Stewart:2010tn}.

A crucial assumption for the derivation of the factorization formula is the absence of Glauber modes~\cite{Stewart:2010tn}, essentially enforcing strict (time-like) soft-collinear factorization from the beginning. Consequently, the result takes the simple form~\eqref{eq:fact_old}, in which the beam and jet functions are color-singlet objects, and the ultra-soft function only contains $(N+2)$ different Wilson lines for the parent partons in the hard scattering process. However, the findings of \cite{Catani:2011st} show that one must carefully distinguish time-like and space-like collinear splittings. The ultra-soft radiation of the beam particles cannot be described by the Wilson line of the incoming parent parton and gives rise to separate outgoing Wilson lines. Moreover, the presence of Glauber gluons entangles the dynamics of the different sectors, such that collinear degrees of freedom remain relevant at the ultra-soft scale.

\begin{figure}
    \centering
    \includegraphics[width=0.8\linewidth]{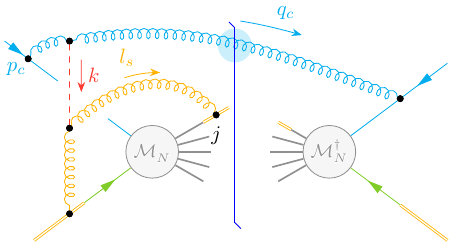} \\[5pt]
    \includegraphics[width=0.8\linewidth]{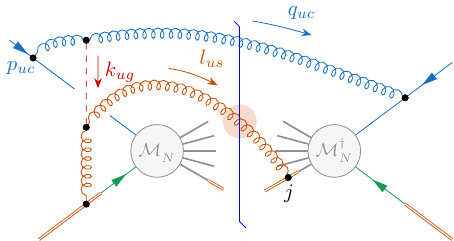}
    \caption{Examples of factorization-breaking diagrams at the collinear (top) and ultra-soft (bottom) scales. A space-like collinear splitting (blue) connects to soft  modes (orange) through Glauber-gluon exchange (red dashed). A colored blob indicates a measurement on the respective line, and $\mathcal{M}_N$ denotes the hard scattering amplitude.}
    \label{fig:coll}
\end{figure}

In the following, we demonstrate the breakdown of the factorization formula~\eqref{eq:fact_old} when these subtleties are taken into account. Due to Glauber interactions, new modes appear in the beam, jet and ultra-soft functions, spoiling naive soft-collinear factorization. The underlying mechanism is similar to the appearance of genuine Glauber modes in the context of non-global observables at three-loop order~\cite{Becher:2024kmk,Becher:2025igg}. For the calculation of Glauber graphs, one can use the SCET Glauber Lagrangian~\cite{Rothstein:2016bsq}.

To be concrete, consider first the beam and jet physics, whose natural scale is $(\mathcal{T}_N\spac Q)^{1/2}$. We introduce the  scaling $(n\cdot p_c,\bar n\cdot p_c,p_{c\perp})\equiv(p_c^+,p_c^-,p_{c\perp})\sim Q\spac(\lambda^2,1,\lambda)$ for the incoming collinear momentum, $p_{\bar{c}}\sim Q\spac(1,\lambda^2,\lambda)$ for the incoming anti-collinear momentum, and likewise for the outgoing jet momenta with respective reference vectors $n_i$. All of these modes were included in the original factorization formula and can both appear virtually and as real emissions which are separately measured.

The presence of Glauber modes in the collinear sector with scaling $k\sim Q\spac(\lambda^2,\lambda,\lambda)$ (and likewise in the anti-collinear one) necessitates the introduction of an additional soft mode characterized by $l_s\sim Q\spac(\lambda,\lambda,\lambda)$. Both modes are purely virtual: the Glauber is off-shell by construction, while real soft emissions are forbidden by the $N$-jettiness measurement function. Since purely virtual contributions are scaleless, these new modes start contributing at three-loop order; an example diagram is shown in the upper part of Fig.~\ref{fig:coll}. Information about the measurement acting on the collinear leg is transferred via the Glauber gluon to the soft loop, where it provides a physical scale. As this is a genuine Glauber mode, well-defined in dimensional regularization and not absorbable into other modes, it needs to be included in our analysis.

\begin{figure}
    \includegraphics[width=0.8\linewidth]{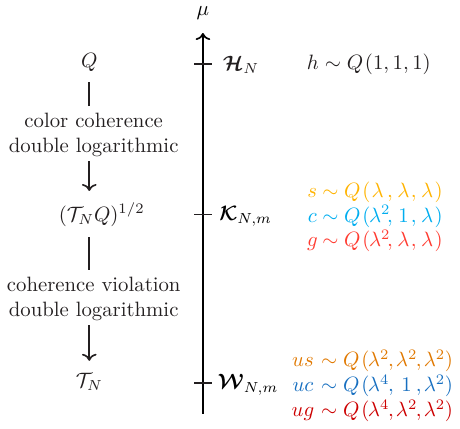}
    \caption{Scale hierarchy and mode structure for $N$-jettiness, color coded for their appearance in diagrams. The soft modes at the collinear scale $(\mathcal{T}_N\spac Q)^{1/2}$ may only appear as virtual particles, while the ultra-collinear modes at the ultra-soft scale $\mathcal{T}_N$ do not contribute to the measurement.}
    \label{fig:hierarchy}
\end{figure}

The physics at the ultra-soft scale $\mathcal{T}_N$ is also affected by Glauber dynamics. This requires the introduction of further modes: alongside the original ultra-soft mode scaling as $l_{us}\sim Q\spac(\lambda^2,\lambda^2,\lambda^2)$, we must include ultra-Glauber modes characterized by $k_{ug}\sim Q\spac(\lambda^4,\lambda^2,\lambda^2)$ and ultra-collinear modes scaling as $q_{uc}\sim Q\spac(\lambda^4,1,\lambda^2)$ (and likewise in the anti-collinear sector). Here, only the ultra-soft mode contributes to the measurement, as the relevant momentum component of ultra-collinear modes is too small, $q_{uc}^+\sim\lambda^4\spac Q$. At the ultra-soft scale, we find similar diagrams as before with the measurement performed on an ultra-soft emission, see the lower graph in Fig.~\ref{fig:coll}. A detailed study of the appearance of genuine Glauber modes for such diagrams was performed in~\cite{Becher:2024kmk,Becher:2025igg} for a slightly different measurement on the soft emission, which does not change the analysis. In addition, it was shown that such diagrams feature a collinear anomaly, resulting in a sensitivity to the hard scale in the form of a rapidity logarithm $\ln(Q/\mu)$ appearing in the low-energy matrix element \cite{Becher:2010tm}, which is untypical for global SCET$_{\mathrm{I}}$ observables. A summary of all modes and the scales at which they appear is shown in Fig.~\ref{fig:hierarchy}.

As a result of the breaking of color coherence and the appearance of these additional modes, the factorization theorem becomes significantly more complicated. Working in Laplace space to turn convolutions into products, and denoting the Laplace variable conjugate to $\mathcal{T}_N$ by $\omega$ (for details see e.g.~\cite{Alioli:2023rxx}), we obtain 
\begin{equation}\label{eq:fact_new}
    \frac{d\sigma}{d\omega} \!=\! \sum_{m} \big\langle \bm{\mathcal{H}}_N(Q,\mu)\,\widetilde{\bm{\mathcal{K}}}_{N,m}(\sqrt{Q\spac\omega},\mu)\,
    \widetilde{\bm{\mathcal{W}}}_{N,m}(\omega,\mu) \big\rangle \,,
\end{equation}
where Laplace-transformed quantities are denoted by tildes. Compared to the naive factorization formula \eqref{eq:fact_old} the hard function remains unchanged, but the inclusion of Glauber modes has drastic consequences for the remaining pieces. This formula takes into account that the final state contains bundles $s_k$ of collinear particles in the beam ($k=a,b$) and jet ($k=1,\dots, N$) directions, labeled collectively by the index $m$. Due to the breaking of color coherence, these can no longer be represented by their parent partons. The quantities $\widetilde{\bm{\mathcal{K}}}_{N,m}$ describe the dynamics of the beam and jet functions, which become entangled beyond two-loop order due to Glauber effects. The low-energy matrix elements $\widetilde{\bm{\mathcal{W}}}_{N,m}$ contain the ultra-soft Wilson lines for the in- and outgoing particles, but also ultra-collinear fields for the initial-state partons. Their explicit form is similar to the low-energy matrix elements for non-global jet observables studied in~\cite{Becher:2025igg}. In analogy with the situation in that paper, PDF factorization becomes non-trivial under these circumstances, and we refrain from factorizing the physics at the ultra-soft scale from non-perturbative dynamics at $\Lambda_{\textrm{QCD}}$. However, due to the structural similarity of the low-energy matrix elements, we strongly believe that PDF factorization is recovered below the ultra-soft scale by the mechanism discovered in~\cite{Becher:2024kmk,Becher:2025igg}.

Another subtlety is that the usual association of \emph{collinear} Wilson lines with the incoming particles no longer holds in this form, and a generalization is required. This is further discussed in the End Matter and connects to the calculation of one-loop space-like splitting functions. Also, beyond two-loop order, it becomes important to carefully keep track of the polarization indices of the collinear fields \cite{Becher:2025igg}.

We now turn to the resummation of large logarithmic corrections in the factorization theorem \eqref{eq:fact_new}, which is sensitive to the appearance of additional modes already in the one-loop anomalous dimension. For concreteness, we focus on the example of $1$-jettiness. The hard function is not modified with respect to the standard treatment, and its anomalous dimension reads (focusing on the cusp logarithm, with $s_{ij}=2q_i\cdot q_j$, and the Glauber phase)~\cite{Becher:2021zkk}
\begin{align}
   \bm{\Gamma}^{H}
    \ni\frac{\alpha_s}{2\pi} \sum_{i\nparallel j}\!
    \left( \bm{T}_i^L\cdot\bm{T}_j^L+\bm{T}_i^R\cdot\bm{T}_j^R \right) 
    \ln\frac{s_{ij}}{\mu^2} + \frac{\alpha_s}{\pi}\,\bm{V}^G \,.
\end{align}
The notation $i \nparallel j$ enforces that parton $j$ is not collinear to parton $i$. The Glauber part relevant for the lowest-order CVL reads \cite{Becher:2023mtx}
\begin{equation}\label{eq:Glauber}
    \bm{V}^G
    = - 2i\pi\,\big( \bm{T}_{1}^L\cdot\bm{T}_{2}^L - \bm{T}_{1}^R\cdot\bm{T}_{2}^R \big) \,.
\end{equation}
The general expression is given in the End Matter. The hard anomalous dimension is purely virtual, as the $N$-jettiness measurement prevents any hard emissions. The labels $L$ and $R$ on color generators indicate whether they act from the left or the right on the hard function, i.e.\ on the amplitude or its conjugate. For the case of $1$-jettiness, the hard anomalous dimension is color-diagonal and the evolution of the hard function to the collinear scale $\omega_c \equiv \sqrt{Q\spac \omega}$ is thus straightforward. 

Below the collinear scale, real emissions are kinematically allowed, and the product  $\bm{\mathcal{H}}_N\spac\widetilde{\bm{\mathcal{K}}}_{N,m}$ evolves to the ultra-soft scale via
\begin{align}\label{eq:RGevol}
   & (\bm{\mathcal{H}}_N\spac\widetilde{\bm{\mathcal{K}}}_{N,m})(\omega) \\
   &= \!\sum_{m'\leq m}(\bm{\mathcal{H}}_N\spac\widetilde{\bm{\mathcal{K}}}_{N,m'} )(\omega_c)\,
    {\bf{P}}\exp\biggl[-\!\int_{\omega}^{\omega_c }\frac{d\mu}{\mu}\,
    \bm{\Gamma}^{\mathcal{W}} \biggr]_{m'm} \,, \nonumber
\end{align}
where $\bm{\Gamma}^{\mathcal{W}}$ denotes the anomalous dimension of the low-energy matrix elements $\widetilde{\bm{\mathcal{W}}}_{N,m}$. For its derivation we introduce a gluon mass at the ultra-soft scale $m_g^2\sim\lambda^4 Q^2$ as an infrared regulator. With the regulator in place, the ultra-collinear modes become visible already at one-loop order. The gluon-mass dependence cancels between the ultra-soft and ultra-collinear modes, leaving behind a large rapidity logarithm. The virtual part of the anomalous dimension matches the hard anomalous dimension, such that $\bm{\Gamma}_{\mathrm{virt}}^{\mathcal{W}}=-\bm{\Gamma}^H$. For the cusp terms in the real-emission contribution we find
\begin{align}\label{eq:SoftRealAD}
   \bm{\Gamma}_{\mathrm{real}}^{\mathcal{W}}
   &\ni \frac{\alpha_s}{\pi} \sum_{i\nparallel j} \bm{T}_i^L\circ\bm{T}_j^R
    \left( \ln\frac{s_{ij}}{\mu^2} - \ln\frac{\mu^2}{\omega^2} - \ln\hat s_{ij} \right) ,
\end{align}
where $\hat s_{ij}=s_{ij}/(Q_i\spac Q_j)$, and the hard scales $Q_i$ are the same for all particles in a given collinear sector. The $\circ$ symbolizes an extension of the color space~\cite{Becher:2021zkk} due to the emission of a gluon collinear to the beams or final-state jets, which raises the multiplicity $m$ by one unit. The low-energy  anomalous dimension is thus not diagonal, but an operator in both color and multiplicity space. The first logarithm in \eqref{eq:SoftRealAD} is due to the collinear anomaly \cite{Becher:2010tm}, whereas the remaining two logarithms are part of the ``naive'' anomalous dimension of the soft function \cite{Jouttenus:2011wh}. In addition to the cusp terms, the anomalous dimension contains color-aware DGLAP terms, see~\cite{Becher:2023mtx,Becher:2025igg}.

Combining the hard and ultra-soft anomalous dimensions, we can obtain the anomalous dimension of the collinear matrix element comprising the dynamics of the beam and jet functions. Using color conservation, the cusp terms in the result $\bm{\Gamma}^{\mathcal{K}}=-\bm{\Gamma}^H-\bm{\Gamma}^{\mathcal{W}}$ can be written in the form
\begin{align}      
   \bm{\Gamma}^{\mathcal{K}}
   \ni \frac{2\alpha_s}{\pi} 
    \sum_{k} \bm{T}_{s_k}^L\circ\bm{T}_{s_k}^R\spac
    \ln\frac{Q_k\spac\omega}{\mu^2} \,; \quad \bm{T}_{s_k}\equiv\sum_{i\in s_k} \bm{T}_i \,,
\end{align}
where the sum runs over $k=a,b,1,\dots,N$, and $s_k$ denotes the set of $k$-collinear partons. Naively, the objects $\bm{T}_{s_k}$ yield the color generators of the parent partons. The above expression generalizes the sum of the one-loop beam and jet anomalous dimensions encountered in the study of the traditional factorization formula \eqref{eq:fact_old} \cite{Jouttenus:2011wh,Stewart:2010qs}. For the cusp terms in the anomalous dimension of the beam functions, in particular, our result implies the replacement $ C_a \to \bm{T}_{s_a}^L\circ\bm{T}_{s_a}^R$ and similarly for $C_b$. Ultimately, this replacement is the source of the coherence violation. Starting at three-loop order, diagrams such as the top graph in Fig.~\ref{fig:coll} interconnect the individual beams and final-state jets, so that $\bm{\Gamma}^{\mathcal{K}}$ can no longer be expressed as a sum over beam and jet anomalous dimensions.

We now proceed with the derivation of the lowest-order CVL for 1-jettiness. We neglect the running of the hard function between the hard and beam scales and use the tree-level expressions $\widetilde{\bm{\mathcal{K}}}_m(\omega_c)=1$ and  $\widetilde{\bm{\mathcal{W}}}_m(\omega)=f_{a/p}(x_1)\, f_{b/p}(x_2)$ for the matching conditions at the collinear and ultra-soft scales, where $f_{a/p}(x)$ denote the proton PDFs. The path-ordered exponential in \eqref{eq:RGevol} leads to a tower of anomalous dimensions acting on the Born-level hard function. To derive the expression for the lowest-order CVL, we use that the first insertion of the anomalous dimension needs to extend the three-parton color space via a real emission, so as to obtain a non-trivial color structure. The next two insertions consist of Glauber operators. They do not commute with the real emission, and an even number of Glauber phases is needed in the (real-valued) cross section. As the fourth operator, we insert the logarithmically-enhanced cusp terms of $\bm{\Gamma}^{\mathcal{W}}$. Under the trace the real and virtual parts combine and the anomaly logarithms cancel, leaving behind a logarithm of the ultra-soft scale. Suppressing the PDFs for brevity, we obtain 
\begin{align}\label{eq:CVL1jet}
   &\frac{d\sigma_{\textrm{CVL}}}{d\omega}
    = \int_\omega^{\omega_c}\!\frac{d\mu_1}{\mu_1} 
    \int_\omega^{\mu_1}\!\frac{d\mu_2}{\mu_2} 
    \int_\omega^{\mu_2}\!\frac{d\mu_3}{\mu_3}
    \int_\omega^{\mu_3}\!\frac{d\mu_4}{\mu_4} \\
   &\quad\times \big\langle\spac\bm{\mathcal{H}}_1^{\mathrm{Born}}\,
    \bm{\Gamma}^{\mathcal{W}}_{\textrm{real}}(\mu_1)\spac
    \frac{\alpha_s(\mu_2)}{\pi}\spac\bm{V}^G\spac 
    \frac{\alpha_s(\mu_3)}{\pi}\spac\bm{V}^G\spac
    \bm{\Gamma}^{\mathcal{W}}(\mu_4)\spac\big\rangle \,, \nonumber
\end{align}
which is non-vanishing since $[\bm{V}^G\!,\bm{\Gamma}^{\mathcal{W}} ]\neq 0$. As the trace vanishes when $\bm{V}^G$ appears at the end, one can also rewrite the last three operators in the form of a double commutator $[\bm{V}^G\!,[\bm{V}^G\!,\bm{\Gamma}^{\mathcal{W}}]]$. Neglecting the running of $\alpha_s$, the iterated scale integral evaluates to
\begin{align}
   &\int_\omega^{\omega_c}\!\frac{d\mu_1}{\mu_1} \ln\frac{\omega_c}{\mu_1}
    \int_\omega^{\mu_1}\!\frac{d\mu_2}{\mu_2} \int_\omega^{\mu_2}\!\frac{d\mu_3}{\mu_3}
    \int_\omega^{\mu_3}\!\frac{d\mu_4}{\mu_4} \ln\frac{\mu_4}{\omega}\nonumber \\[1mm] 
   &= \frac{1}{46080}\spac\ln^6\!\bigg(\frac{Q}{\omega} \bigg) \,.
\end{align}
For the $qg\to q$ channel, expression \eqref{eq:CVL1jet} confirms the corresponding result obtained in~\cite{Banfi:2025mra}. Our formalism also makes it clear that the leading CVLs scale as $d\sigma_{\textrm{CVL}}/d\omega\sim\alpha_s^{2+n}\ln^{2+2n}(Q/\omega)$ with $n\geq 2$. For $0$-jettiness, the CVL series starts at five-loop order, because an additional insertion of $\bm{\Gamma}^{\mathcal{W}}_{\textrm{real}}$ is needed for a non-trivial color structure.

Formally, the CVLs constitute a leading double-logarithmic contribution to the cross section which does not exponentiate in a simple form. Similar to SLLs in non-global observables, CVLs are suppressed in both color and higher loop order. Yet, as shown in~\cite{Becher:2024nqc} for SLLs, these effects can still be significant and affect precision studies of global observables. 

It seems remarkable that the complicated structure of the factorization theorem \eqref{eq:fact_new} and the presence of several additional modes have not been observed previously. Despite the fact that computations of the beam, jet and ultra-soft functions have been carried out up to three-loop order~\cite{Bruser:2018rad,Banerjee:2018ozf,Ebert:2020unb,Baranowski:2022vcn,Baranowski:2024vxg}, with resummation achieved to N$^3$LL accuracy~\cite{Alioli:2023rxx}, no inconsistencies were encountered. The reason is simple yet concerning: a factorization formula assuming time-like kinematics is internally self-consistent. By working exclusively within such a framework, and neglecting Glauber effects in all components of the factorization formula, the additional contributions remain entirely hidden, regardless of the loop order at which the calculations are performed. The only possibility to detect inconsistencies would be the direct QCD computation of the hard function in physical kinematics, with inconsistencies starting at four-loop order for pure QCD processes~\cite{Forshaw:2012bi}. We expect that three-loop effects stemming from $\bm{\mathcal{H}}_N\bm{V}^G\bm{V}^G\spac\bm{\Gamma}^{\mathcal{W}}$ under the trace in \eqref{eq:CVL1jet} are captured by the naive factorization formula \eqref{eq:fact_old}.

In this Letter, we have uncovered several key ingredients for the correct resummation of global hadron-collider observables at the example of $N$-jettiness. The presence of Glauber modes gives rise to a significantly more involved factorization theorem: Glauber exchanges break strict factorization and introduce several additional modes, resulting in a richer and more intricate structure. These changes will similarly affect other global observables, where space-like splittings are allowed and which are hadronically not fully inclusive. For example, we expect that jet-veto resummation~\cite{Banfi:2012yh,Becher:2012qa,Banfi:2012jm} as well as energy-energy correlators~\cite{Moult:2025nhu} are affected, while $q_T$ resummation is not, as only electroweak bosons are measured in the final state. We have presented a framework for the resummation of coherence-violating logarithms to higher loop orders, the full understanding of which will be crucial going forward. Our findings underscore the urgency of further developing finite-$N_c$ parton showers~\cite{Nagy:2007ty,Nagy:2019pjp,Forshaw:2019ver,DeAngelis:2020rvq,Forshaw:2025bmo,Forshaw:2025fif} and achieving a deeper understanding of the underlying mode structure in general~\cite{Beneke:1997zp,Jantzen:2012mw,Gardi:2024axt,Jones:2024mfg,Ma:2026pjx}, verifying that no further modes enter the factorization theorem in higher loop orders. The generalization of the collinear Wilson-line structures discussed in the End Matter and diagrams such as the first graph in Fig.~\ref{fig:coll} form the basis for the calculation of the factorization-breaking terms in space-like splitting functions. We verified that this approach yields the correct one-loop factorization-breaking terms; the two-loop calculation is left for future work. 

The formalism developed here will have a direct impact on LHC precision studies utilizing $N$-jettiness or other global observables, including Higgs and electroweak physics as well as dark matter searches. They also demonstrate that the structure of SCET factorization for hadron colliders is much richer than previously appreciated. Indeed, the additional complexity should be viewed as an opportunity rather than an obstacle, as it opens new avenues for theoretical exploration, phenomenological investigation, and precision tests of QCD.

\begin{acknowledgments}
{\em Acknowledgements ---} 
We thank Andrea Banfi for triggering this project during a workshop at the Mainz Institute of Theoretical Physics (MITP). 
We are grateful to Jeff Forshaw, Einan Gardi, and Jack Holguin for useful discussions. TB and DS would like to thank the MITP for hospitality and support. This work was supported by the Swiss National Science Foundation (SNSF) under grant 200021\_\spac 219377, the European Research Council (ERC) under the European Union’s Horizon 2022 Research and Innovation Program (ERC Advanced Grant agreement No.~101097780, EFT4jets), and the Cluster of Excellence PRISMA$^{++}$ (Precision Physics, Fundamental Interactions, and Structure of Matter, EXC 2118/2) funded by the German Research Foundation (DFG) under Germany’s Excellence Strategy (Project ID 390831469).
\end{acknowledgments}

\bibliography{cvl}

\section{End matter}

\noindent\textbf{General form of the Glauber operator }\vspace{-2mm}\\ 

The general expression for the one-loop Glauber operator is given by \cite{Becher:2021zkk}
\begin{align}
   \bm{V}^G = - \frac{i\pi}{2} \sum_{i\nparallel j}\Pi_{ij} \bigl( \bm{T}_{i}^L\cdot\bm{T}_{j}^L-\bm{T}_{i}^R\cdot\bm{T}_{j}^R \bigr) \,,
\end{align}
with $\Pi_{ij}=1$ if $i$ and $j$ are both either ingoing or outgoing, and $\Pi_{ij}=0$ otherwise. Using color conservation, this expression can be rewritten in the form
\begin{align}
   \bm{V}^G 
   &= - \frac{i\pi}{2}\,\bigg[\spac 4\spac\bm{T}_a^L\cdot\bm{T}_b^L 
    - \left( \bm{T}_{s_a}^L - \bm{T}_a^L \right)^2 - \left( \bm{T}_{s_b}^L - \bm{T}_b^L \right)^2 \nonumber\\
   &\hspace{1.3cm} - \sum_{k=1}^N \left( \bm{T}_{s_k}^L \right)^2 - (L\to R) \bigg] \,.
\end{align}
For the calculation of the leading-order CVLs the extra terms compared with~\eqref{eq:Glauber} reduce to Casimir operator and cancel out.
\vspace{4mm}

\noindent\textbf{Collinear Wilson lines}\vspace{-2mm}\\ 

At the intermediate scale $(\mathcal{T}_N\spac Q)^{1/2}$, the SCET building blocks contain collinear Wilson lines. However, when space-like splittings are involved, the usual expression for these Wilson lines needs to be generalized. For an amplitude with only incoming particles, attaching an $n_i$-collinear gluon to a leg $j\nparallel i$ produces an off-shell hard fluctuation, which can be integrated out to give the Wilson line
\begin{equation}
   \bm{W}_{i,j}^{+}(x) = {\bf{P}}\exp\!\bigg[ -i g_s\int_{-\infty}^0\!ds\, \bar n_i\cdot A_i^{a}(x+s\spac\bar n_i)\,\bm{T}_j^{a} \bigg] \,,
\end{equation}
where ${A}_i^{\mu,a}$ is the associated $n_i$-collinear gluon field. The path from $-\infty$ to $0$ corresponds to using the $+i0$ prescription in the eikonal propagator with incoming momentum $k_i$. Combining the attachments to all legs $j\nparallel i$ yields
\begin{align}\label{eq:Wilsonproduct}
  {\bm{W}}_i(x)  & 
   = {\bf{P}}\exp\!\bigg[-i g_s\int_{-\infty}^0\!ds\,
    \bar n_i\cdot A_i^{a}(x+s\spac\bar n_i)\sum_{j\neq i}\bm{T}_j^{a} \bigg] \nonumber\\
   &= {\bf{P}} \exp\!\bigg[ \spac i g_s\int_{-\infty}^0\!ds\,
      \bar n_i\cdot A_i^{a}(x+s\spac\bar n_i)\,\bm{T}_i^{a} \bigg]
    \,,
\end{align}
where we have used color conservation ($\sum_i\bm{T}_i=0$) in the last step. The result is the familiar form of the collinear Wilson line in SCET. This form allows for the factorization of the different collinear sectors from each other, but crucially assumes the same $i0$ prescription for all collinear propagators (all particles incoming). Typically, it is argued that the $i0$ prescription in the eikonal propagators is irrelevant, which is indeed true for purely time-like kinematics.

\begin{figure}
    \centering
    \includegraphics[width=0.8\linewidth]{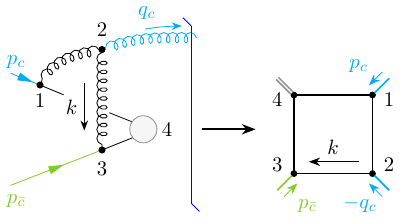}
    \caption{Mapping of the one-loop space-like splitting function onto a box diagram. The external momentum of the double line is off-shell, $(p_c-q_c+p_{\bar{c}})^2\neq 0$.}
    \label{fig:box}
\end{figure}

In the following, we show explicitly that for space-like collinear splittings, the $i0$ prescription does matter and that the usual Wilson line needs to be generalized. We start with the calculation of the diagram shown in Fig.~\ref{fig:box}, where only the scalings of the in- and outgoing collinear momenta $p_c,q_c$ and anti-collinear momentum $p_{\bar{c}}$ are fixed, while the scaling of the loop momentum $k$ is left unspecified. We strip off the Lorentz structure and map this diagram onto the box graph shown on the right. We can then ascertain that all leading-order scalings of the virtual momentum $k$ are accounted for by comparing the method-of-region result to the well-known result for the box diagram, expanded to leading order in $\lambda$~\cite{Bern:1993kr}. We find that the leading contributions arise from collinear or Glauber scaling. However, the Glauber region is scaleless after analytic regularization. The collinear region gives rise to the integral
\begin{align}\label{eq:collInt}
   J &= - i\spac(4\pi)^{d/2} \int\!\frac{d^dk}{(2\pi)^d}\spac
    \frac{1}{k^2+i0}\spac\frac{1}{(k+q_c)^2+i0} \nonumber\\ 
   &\quad\times \frac{1}{(k+q_c-p_c)^2+i0}\spac\frac{1}{p_{\bar{c}}^+k_-+i0} \,. 
\end{align}
Performing the integrals yields
\begin{align}
    J &= \frac{\Gamma^2(\varepsilon)\spac\Gamma^2(1-\varepsilon)}{s_{12}\spac s_{23}\spac s_{45}\spac\Gamma(-2\epsilon)} 
     \left( -s_{12} \right)^{-\epsilon} \bigg[\spac s_{45} \left( -1 + \frac{s_{45}}{s_{23}} \right)^\epsilon \Gamma(-\epsilon) \nonumber\\
    &\quad - \frac{s_{23}}{\Gamma(2+\epsilon)}\,{}_{2}F_{1}\!\left(1,1,2+\epsilon,\frac{s_{23}}{s_{45}}\right) \bigg] \,,
\end{align}
with $s_{12}=-p_c^-q_c^+$, $s_{23}=-p_{\bar{c}}^+q_c^-$ and $s_{45}=p_{\bar{c}}^+(p_c^--q_c^-)$. Comparison with the result~\cite{Bern:1993kr} shows perfect agreement in Euclidean kinematics, where all $s_{ij}<0$. Performing the analytic continuation $s_{ij} \to s_{ij}+i0$ for both time-like and space-like splittings, we find that space-like splittings are sensitive to the $i0$ prescription while time-like splittings are not. As the eikonal propagator in~\eqref{eq:collInt} corresponds to the collinear Wilson line, it is clear that the $i0$ prescription matters. This demonstrates that the simple Wilson line in \eqref{eq:Wilsonproduct}, where one is indifferent towards the $i0$ prescription, is insufficient.

Indeed, distinguishing the Wilson lines for incoming and outgoing particles carefully, we find the proper form
\begin{equation}
   \prod_{j\neq i} \bm{W}_{i,j}^{+}(x)
   \to \prod_{k\neq i} \bm{W}^+_{i,k}(x) \prod_{l\neq i} \bm{W}^-_{i,l}(x) \,,
\end{equation}
where $k$ refers to incoming and $l$ to outgoing legs, and we have defined
\begin{equation}
   \bm{W}_{i,j}^{-}(x) = \overline{\bf{P}}\exp\!\bigg[\spac i g_s\int_0^{\infty}\!ds\, \bar n_i\cdot A_i^{a}(x+s\spac\bar n_i)\,\bm{T}_j^{a} \bigg] \,.
\end{equation}
We can combine the Wilson lines by introducing unit operators $\bm{1}=\bm{W}_{i,l}^+\,(\bm{W}_{i,l}^+)^\dagger$, which leads to
\begin{align}
   \bm{W}_{i}(x) 
   \to \bm{W}_{i}(x) \prod_{l\neq i}\,(\bm{W}^+_{i,l})^\dagger(x)\,\bm{W}^-_{i,l}(x) \,.
\end{align}
The extra terms, which arise from changing the $i0$ prescription for incoming legs, give rise to $\delta$-distributions. Notably, this effect occurs only for processes  with both in- and outgoing particles and a space-like collinear emission on top of it, in agreement with the discussion in~\cite{Catani:2011st}. We believe that this effect is exclusive to space-like splittings of initial-state fields; the modifications of the Wilson lines are of no consequence for outgoing collinear fields. We have moreover confirmed that the factorization-breaking terms of the one-loop space-like splitting function can be completely reproduced using this prescription.

We emphasize that this kind of factorization breaking is not driven by a (genuine) Glauber mode or region, but can be accounted for by a non-vanishing Glauber limit of the collinear integral, which shows itself through sensitivity to the $i0$ prescription, see the last propagator in the integral in~\eqref{eq:collInt}. Working in light-cone gauge, where $\bm{W}_i^+(x)=1$, does not change this fact, as the $i0$ prescription in the light-cone propagator must then be kept accordingly.
\vfil

\end{document}